\documentclass[twoside]{article}
\usepackage{fleqn,espcrc2}

\newcommand{\AmS}{{\protect\the\textfont2
  A\kern-.1667em\lower.5ex\hbox{M}\kern-.125emS}}

\title{Reformulating Yang-Mills theory in terms of local gauge invariant variables}

\author{Pushan Majumdar \address[IMSC]{Institute of Mathematical Sciences,
        C.I.T. Campus Taramani, Chennai - 600113. India.}%
	\thanks{Present address: Dept. of Theoretical Physics, Tata Institute
	of Fundamental Research, Homi Bhabha Road, Mumbai 400005, India.}
	and
        H.S.Sharatchandra \addressmark[IMSC]} 
       
\begin{document}

\begin{abstract}
An explicit canonical transformation is constructed to relate the physical 
subspace of Yang-Mills theory to the phase space of the ADM variables of general 
relativity. This maps 3+1 dimensional Yang-Mills theory to local evolution of 
metrics on 3 manifolds. 
\vspace{1pc}
\end{abstract}

\maketitle

%\section{Introduction}

A long standing problem in Yang-Mills theories is whether its 
dynamics can be written in terms of gauge invariant
variables. This is very important for our understanding of 
confinement in QCD. 
One option is to write the theory as dynamics of
Wilson loops. However these are non-local variables and their evolution 
equations are difficult to handle. Another approach has been to use 
composite variables constructed out of the non-Abelian electric \cite{1,2,2a}
or magnetic fields \cite{4,5}.  

In three dimensions, gauge invariant formulation of Yang-Mills theory 
was carried out in close analogy to gravity in \cite{9}. A similar approach 
was tried out for 3+1 dimensions in \cite{10}, but the theory turned out to 
be quite complicated. Here we do a manifestly gauge invariant 
formulation of 3+1 dimensional Yang-Mills theory using Ashtekar like 
variables. This formulation is also closely related to ADM formulation 
of gravity. We identify the physical variables and the Gauss law and  
then rewrite the Hamiltonian in terms of the new variables. 

Starting with the usual Euclidean partition function, we introduce an 
auxiliary field $E^{ia}$ and integrate over $A_0^a$ to obtain
\begin{eqnarray}\label{fgauss}
Z&=&\int\:{\cal D} A_{i}^{a}{\cal D} E^{ia}\;\delta (D_{i}[A]E^{i})\;\nonumber \\
 &&exp \{ \int (-{\cal H}+i\vec{E}^{i}\cdot\partial_{0}\vec{A}_{i})\}.
\end{eqnarray}
The hamiltonian density ${\cal H}$ is given by
\begin{equation}\label{ham}
{\cal H}=\frac{1}{2} (g^{2}E^{i2}+\frac{1}{g^{2}}B_i^{2}),
\end{equation}
where $B_i^a$ is the usual non-Abelian magnetic field.
 The phase space of the theory is defined in terms of the 
canonically conjugate variables $A_i^a$ and
$E_i^a$. These variables are not free, but
satisfy a constraint equation which is the non-Abelian Gauss law
\begin{equation}\label{1}
\partial_i{\vec E}^i+{\vec A}_i\times {\vec E}^i =0.
\end{equation}

Following Ashtekar \cite{7} let us define the driebein $e$ as
\begin{equation}\label{2}
{\vec E}^i=\frac{1}{2}\epsilon^{ijk}{\vec e}_j\times {\vec e}_k
\end{equation}
Assuming $||E||=||e||^2$ is nonzero, we obtain
$ e_i^{a} = ||E||^{\frac{1}{2}}(E^{-1})_i^{a}$.
Next we introduce the connection ${\vec {\tilde A}}_i(E)$ such that
\begin{equation}\label{3}
\epsilon_{ijk}(\partial_j e_k + {\vec {\tilde A}}_j(E)\times {\vec e}_k)=0   
\end{equation}
This is the unique connection one form which is torsion free with respect 
to the driebein $e$. ${\tilde A}_i^a$ and $E_i^a$ together satisfies the 
non-Abelian Gauss law as follows from (\ref{1}) and (\ref{2}).
Let us denote the difference between ${\vec A}_i$ and ${\vec {\tilde A}}_i$ by ${\vec a}_i$.
In terms of the new variable ${\vec a}_i$ the Gauss law can be written as 
\begin{equation}\label{4}
{\vec a}_i\times {\vec E}^i=0
\end{equation}
Note that since ${\vec a}_i$ is the difference between two connections, it transforms 
homogenously under gauge transformations.

The transformation from $\{A_i, E^i\}$ to $\{a_i,E^i\}$ is a canonical
transformation generated by \\
\begin{equation}\label{5}
S[a_i^a, E^{ia}]=\int d^3x\;a_i^aE^{ia}+\frac{1}{2}\epsilon^{ijk}{\vec
e}_i\cdot\partial_j{\vec e}_k
\end{equation}
We now want to make a variable transformation from the pair $\{a_i^a, E_i^a\}$ to the
pair $\{\pi^{ij}, g_{ij}\}$. We will again do it by a canonical transformation.
The main reason for using canonical transformations is that this leaves the
phase space measure in the
functional integral invariant in terms of the new variables.
As the variables
$(a,E)$ are more in number than ($\pi, g$), we first augment the latter set by a
canonically conjugate set $\{\chi^a, \theta_a\}$.
The relation between the old momenta and new coordinates are
\begin{equation}\label{9}
g_{ij}={\vec e}_i\cdot {\vec e}_j.
\end{equation}
To define $\theta_a$ consider the Polar decomposition of $e_i^a$.
\begin{equation}\label{10}
e_i^a=e_{ij}O^{ja}
\end{equation}
where $e_{ij}$ is a symmetric matrix and 
$O^{ja}$ is an orthogonal matrix. 
$\theta^a$ is the lie algebra element corresponding to $O^{ja}$.
\begin{equation}\label{11}
O=exp (i\theta_a \frac{T^a}{2}).
\end{equation}
To construct the full transformation, consider the generator of the canonical transformation 
\begin{equation}\label{5}
S(\pi,\chi,E)=\int \left (({\vec e}_i\cdot {\vec e}_j)\pi^{ij} + \theta_a (e)\chi^a
\right)
\end{equation}
From this generating functional, we have the relations
\begin{eqnarray}
g_{ij}\equiv\frac{\delta S}{\delta \pi^{ij}}&=&{\vec e}_i\cdot {\vec e}_j \\
\theta_a\equiv\frac{\delta S}{\delta \chi^{a}}&=&\theta_a[e]\\
a_i^a\equiv\frac{\delta S}{\delta E^{ia}}&&
\end{eqnarray}
To relate `$a$' explicitly to the other variables, note that
\begin{equation}\label{16}
\frac{\delta S}{\delta e_i^a}=\frac{\delta E^{bj}}{\delta e_i^a}
\frac{\delta S}{\delta E^{bj}}
=\frac{\delta E^{bj}}{\delta e_i^a}a_j^b.
\end{equation}
In terms of these new variables, Gauss law becomes
\begin{equation}\label{18}
({\vec a}_i \times {\vec E}^i)^a=\frac{1}{2}(M^{-1})_b^a[\theta]\chi^b
\end{equation}
where $M$ is defined by
\begin{equation}
O^T(\theta )O(\theta + \delta \theta )\approx 1+T^a M^a_b(\theta)\delta \theta^b.
\end{equation}
It is important to note that the canonical momentum $\pi^{ij}$ drops out in the
Gauss law equation. {\em With the new variables, the Gauss law is implemented by
simply setting $\chi=0$.} 
We have thus achieved a separation between the physical momenta $\pi^{ij}$ and
the constraint momenta $\chi$.
This is in direct analogy to ADM decomposition of the metric in gravity, which 
gives action as 
\begin{equation}
I=\int\,d^4x\,{\sqrt g}R=\int\,d^4x\,\{\pi^{ij}{\dot g}_{ij}-N_{\mu}C^{\mu}\}
\end{equation}
where $\pi^{ij}$ and $g_{ij}$ are the physical variables and $N_{\mu}$ and 
$C^{\mu}$ are the constraints involving the lapse function and shift vectors. 
In terms of the new variables, our partition function becomes 
\begin{eqnarray}\label{20}
Z&=&\int{\cal D}g_{ij}\,{\cal D}\pi^{ij}\,{\cal D}\theta^a\,{\cal D}\chi^a
\delta((M^{-1})^a_b[\theta]\chi^b) \nonumber \\
&&exp\int(-{\cal H}^{'}+i\pi^{ij}\partial_0g_{ij}+i\chi^a\partial_0\theta^a).
\end{eqnarray}
$\theta^a$ represents the gauge degrees of freedom. We may adopt the
Faddeev-Popov
procedure to choose $\theta^a=0$. In this case $M^a_b[\theta]=\delta^a_b$, and
\begin{equation}\label{21}
Z=\int{\cal D}g_{ij}\,{\cal D}\pi^{ij}\,exp\int(-{\cal H}^{'}[g,\pi]+i\pi^{ij}
\partial_0g_{ij})
\end{equation}
Now we have to rewrite the Hamiltonian in terms of the new variables. The 
relations between the old and the new variables are given by
\begin{eqnarray}
E^{ia}&\rightarrow & \frac{1}{2}\epsilon^{ijk}\epsilon^{abc}e_{jb}e_{kc} \label{21}\\
A_i^a&\rightarrow & {\bar A}_i^a[E]+a_i^a \label{22}\\
a_i^a&\rightarrow & \frac{1}{||e||}(\pi^{jk}g_{jk}e_{ia}-2\pi^{jk}g_{ik}e_{ja})\label{23}.
\end{eqnarray}
Here $e_{ia}$ has to be regarded as the symmetric square root of $g_{ij}$.
Hence it has only six degrees of freedom and not the usual nine. From (\ref{21}) 
$({\vec E}^i)^2 $ becomes $ \frac{g^{ii}}{||g||}$.
For $B_i[A]$ we can do an expansion about ${\tilde A}[E]$
\begin{eqnarray}\label{24}
B_i[A] &=& B_i[{\tilde A}[E]]+\epsilon_{ijk}D_j[{\tilde A}[E]]a_k
\nonumber \\ &&+\frac{1}{2} \epsilon_{ijk}(a_j\times a_k).
\end{eqnarray}
$ B_i[{\tilde A}[E]]$ can be expressed entirely in terms of $g_{ij}$ and its 
derivatives. In fact it is related to the Ricci tensor.
\begin{equation}
B_i^a[{\tilde A}[E]]=\frac{1}{4||e||}\epsilon_{ijk}\,\epsilon_{lmn}
R_{jl}\,g_{km}\,e_n^a 
\end{equation}
To evaluate the square of the last term in (\ref{24}), we use 
\begin{eqnarray}
&\epsilon_{ijk}(a_j\times a_k)\cdot\epsilon_{imn}(a_m\times a_n)\hspace*{.5in} &
 \nonumber \\&\hspace*{.5in}=({\vec a}_j\cdot {\vec a}_m)({\vec a}_j
\cdot{\vec a}_m) -({\vec a}_j\cdot{\vec a}_j)^2. &
\end{eqnarray}
Here finally `$a$' has to be replaced by the expression in (\ref{23})

Let us define $\Gamma_{ij}^k[g]$ by the equation
\begin{equation}
(D_i[{\bar A}[E]]e_j)^a=\Gamma_{ij}^k[g]e_k^a.
\end{equation}
Now we can replace $D_k[{\bar A}[E]]$ by the covariant derivative 
corresponding to $\Gamma_{ij}^k[g]$ when acting on $g_{ij}$ or $\pi^{ij}$.
Thus $({\vec B}_i[A])^2$ and hence the full Hamiltonian can be written in 
terms of local gauge invariant quantities.

We have thus mapped the physical phase space of Yang-Mills theory onto the
phase space of the ADM variables and the dynamics is now a local evolution
of the metrics on 3-manifolds.
The constraint variables and physical variables are explicitly separated
and this greatly simplifies the imposition of Gauss law.
The theory is now described in terms of gauge invariant unconstrained
variables.


\begin{thebibliography}{9}
\bibitem{1} R.Anishetty, Phys. Rev. D 44, (1991) 1895.
\bibitem{2} F.A.Lunev, Phys. Lett. B 295, (1992) 99.
\bibitem{2a} A. M. Khvedelidze and H.-P. Pavel, Phys. Rev. D 59, (1999) 105017.
\bibitem{4} Peter E. Haagensen and Kenneth Johnson, Nucl. Phys. B 439, (1995) 597.
\bibitem{5} Peter E. Haagensen, Kenneth Johnson and C.S.Lam, Nucl.Phys. B 477,
(1996) 273.
\bibitem{9} Ramesh Anishetty, Pushan Majumdar and H.S.Sharatchandra, Phys. Lett.
 B 478, (2000) 373
\bibitem{10} F.A.Lunev, Modern Physics Letters A Vol 9 No. 25, (1994) 2281.
\bibitem{7} A.Ashtekar, Phys. Rev. Lett. 57, (1986) 2244.
\end{thebibliography}
\end{document}